# Cluster property and entanglement in the superconducting state

Katsuhiko Higuchi[1*] and Masahiko Higuchi[2*]

[1] *Graduate School of Advanced Sciences of Matter, Hiroshima University, Higashi-Hiroshima 739-8527, Japan*
[2] *Department of Physics, Faculty of Science, Shinshu University, Matsumoto 390-8621, Japan*

* Contact authors: khiguchi@hiroshima-u.ac.jp (K. Higuchi); higuchi@shinshu-u.ac.jp (M. Higuchi)

**Abstract**

We present the conditions for satisfying the cluster property in the superconducting state. Using these conditions, we show that the cluster property employed when deriving the order parameter of the superconducting state (OPSS) holds true, and that the OPSS is given by ${}_S\langle\Psi|\hat{\psi}(x)\hat{\psi}(x')|\Psi\rangle_S$, where $|\Psi\rangle_S$ is the superconducting state, and where $\hat{\psi}(x)$ is the field operator of superconducting electrons. This provides a resolution to the long-standing concern raised by Leggett *et al*. regarding particle-number fluctuation in the superconducting state. In addition, by considering the entanglement between two particles composed of the maximum geminal of the BCS ground state [J. Phys. Commun. **5**, 095003 (2021)], it is confirmed that the larger the deviation from the conditions for satisfying the cluster property becomes, the greater the magnitude of the entanglement becomes. This shows not only the effectiveness of the conditions for satisfying the cluster property but also the justification of the OPSS that has been commonly used.

## I. Introduction

When the system enters the superconducting state, the order parameter of the superconducting state (OPSS) emerges. The following quantity has been widely recognized as the OPSS [1-18]:

$$ {}_S\langle \Psi | \hat{\psi}(\boldsymbol{r},\zeta)\hat{\psi}(\boldsymbol{r}',\zeta') | \Psi \rangle_S , \tag{1} $$

where $|\Psi\rangle_S$ is the superconducting state, and where $\hat{\psi}(\boldsymbol{r},\zeta)$ is the field operator of superconducting electrons, and where $\boldsymbol{r}$ and $\zeta$ are spatial and spin coordinates, respectively.

It has been shown that the OPSS takes the form of Eq. (1) under the assumption that the cluster property holds in the superconducting state [1-5]. The cluster property in the superconducting state can be stated as follows: the second-order reduced density matrix (RDM2) for the superconducting state, which is defined as ${}_S\langle \Psi | \frac{1}{2}\hat{\psi}^\dagger(\boldsymbol{r}_1',\zeta_1')\hat{\psi}^\dagger(\boldsymbol{r}_2',\zeta_2')\hat{\psi}(\boldsymbol{r}_2,\zeta_2)\hat{\psi}(\boldsymbol{r}_1,\zeta_1) | \Psi \rangle_S$, is clustered like $\frac{1}{2}\,{}_S\langle \Psi | \hat{\psi}^\dagger(\boldsymbol{r}_1',\zeta_1')\hat{\psi}^\dagger(\boldsymbol{r}_2',\zeta_2') | \Psi \rangle_S \, {}_S\langle \Psi | \hat{\psi}(\boldsymbol{r}_2,\zeta_2)\hat{\psi}(\boldsymbol{r}_1,\zeta_1) | \Psi \rangle_S$ under the simultaneous limits of $\boldsymbol{r}_1 \approx \boldsymbol{r}_2$, $\boldsymbol{r}_1' \approx \boldsymbol{r}_2'$ and $|\boldsymbol{r}_1 - \boldsymbol{r}_1'| \to \infty$ [1-5]. If this cluster property holds, when the superconducting state is regarded as the Bose-Einstein condensation (BEC) of fermion system [4, 5, 19, 20], the condensing two-particle state and Eq. (1) have the following relation [1, 2, 4]:

$$ {}_S\langle \Psi | \hat{\psi}(\boldsymbol{r}_1,\zeta_1)\hat{\psi}(\boldsymbol{r}_2,\zeta_2) | \Psi \rangle_S = \sqrt{2\, n^{(2)}_{\nu_\Psi^{\max}}}\, \nu_\Psi^{\max}(\boldsymbol{r}_1,\zeta_1,\boldsymbol{r}_2,\zeta_2) , \tag{2} $$

where $n^{(2)}_{\nu_\Psi^{\max}}$ is the occupation number of the condensing two-particle state $\nu_\Psi^{\max}(\boldsymbol{r}_1,\zeta_1,\boldsymbol{r}_2,\zeta_2)$. That is to say, if the cluster property holds in the superconducting state, Eq. (1) can be regarded as the OPSS. Does Eq. (2) hold true? In other words, does the cluster property hold in the superconducting state? Equation (1) definitely shows that the superconducting state is a state in which the particle number is not conserved. Is this also true?

It has been shown in the field of the first-principles calculations for the superconductor that Eq. (1) works well as the OPSS, particularly for the quantitative evaluation of the superconducting transition temperature [2, 10-18]. On the other hand, some workers have doubts about whether the cluster property holds in the superconducting state [21-23]. The basis of this doubt is likely an intuitive discomfort with the fluctuation of the particle number in the system since electrons practically have an infinite lifetime [21-23]. Actually, Leggett proposed a theory of superconductivity using the superconducting state with a fixed number of particles [21-23]. As is well established, the BCS theory has achieved great success for explanations of various superconducting phenomena [24-26].

In the BCS theory, the particle number fluctuates in the ground state. Furthermore, it has been recently shown that the cluster property holds true within the BCS theory, i.e., Eq. (2) holds true within the BCS theory [27]. However, it is not trivial whether the cluster property holds in a general superconducting state. One of the purposes of this paper is to show that above-mentioned doubt is likely unnecessary even for a general superconducting state outside of the BCS framework.

In general, if the state of a composite system cannot be expressed as a single tensor product of the states of the individual systems, the cluster property between two individual systems is broken [28]. Equivalently, it can be said that the entanglement between two individual systems emerges in that case [28]. In this paper, we shall first derive the conditions for satisfying the cluster property between individual two system. This means that we derive the conditions for generating entanglements between two individual systems. The details will be shown in Sec. III.

Next, we will consider these conditions for satisfying the cluster property in the case of superconducting state. Here note that there exist two types of entanglements in the superconducting state. One is the entanglement between two systems both of which consist of the collection of the maximum geminals [27] that correspond to the so-called Cooper pairs of the superconducting state. The other is the entanglement between two particles constituting the maximum geminal itself. In Sec. IV, we will consider the conditions for satisfying the cluster property between two systems composed of the maximum geminals, which is referred to as the former type above. Using these conditions, it can be proven that the cluster property used when deriving the OPSS holds true. This provides a response to the doubt raised by Leggett *et al.* mentioned above [21-23] and, at the same time, shows the validity of the OPSS being in the form of Eq. (1). The details will be shown in Sec. V.

Subsequently, in Sec. VI, we will discuss the entanglement in the superconducting state, which is referred to as the latter type above. In this case, the conditions for satisfying the cluster property are also utilized. The larger the deviation from the conditions for satisfying the cluster property becomes, the larger the magnitude of the entanglement becomes. This also shows the effectiveness of the conditions for satisfying the cluster property in the superconducting state.

## II. Antisymmetrizing operator

We consider two systems named as $S_1$ and $S_2$, and consider a composite system $S$ that consists of system $S_1$ and system $S_2$. There exist $N_1$ and $N_2$ electrons in system $S_1$ and system $S_2$, respectively, and exist electrons $N\ (= N_1 + N_2)$ in a composite system $S$. Suppose that the number of electrons is conserved in system $S_1$ and system $S_2$, respectively. Let the coordinates of electrons in system $S_1$ and system $S_2$ be written as $x_1, x_2, x_3, \cdots, x_{N_1}$, and $y_1, y_2, y_3, \cdots, y_{N_2}$, respectively, where the coordinate $x_i$; $i = 1, 2, 3, \cdots, N_1$ or $y_j$; $j = 1, 2, 3, \cdots, N_2$ contains both spatial coordinates and spin coordinates.

The Hamiltonian of system $S_1$ is written as

$$H^{(1)} = \sum_{j=1}^{N_1} h^{(1)}\left(x_j, \nabla_{x_j}\right) + \frac{1}{2}\sum_{i=1}^{N_1}\sum_{j=1(\neq i)}^{N_1} w^{(1)}\left(x_i, x_j\right), \tag{3}$$

where $h^{(1)}\left(x_j, \nabla_{x_j}\right)$ is the sum of the kinetic and potential energies for the $j$th electron, and $w^{(1)}\left(x_i, x_j\right)$ is the interaction energy between the $i$th and $j$th electrons. Similarly, the Hamiltonian of system $S_2$ is written as

$$H^{(2)} = \sum_{j=1}^{N_2} h^{(2)}\left(y_j, \nabla_{y_j}\right) + \frac{1}{2}\sum_{i=1}^{N_2}\sum_{j=1(\neq i)}^{N_2} w^{(2)}\left(y_i, y_j\right). \tag{4}$$

When the interaction energy between the $i$th electron of system $S_1$ and the $j$th electron of system $S_2$ is written as $W^{(12)}\left(x_i, y_j\right)$, the Hamiltonian of the composite system $S$ is given by

$$H = H^{(1)} + H^{(2)} + W^{(12)}. \tag{5}$$

***Eigenfunctions for $H^{(1)}$ and $H^{(2)}$***

As is well known, the eigenfunction for $H^{(1)}$ is generally given by

$$\Phi_n^{(1)}\left(x_1, x_2, \cdots, x_{N_1}\right) = \sum_i a_{i,n}^{(1)}\left[\frac{1}{\sqrt{N_1!}}\sum_{P\in S_1}(-1)^p P\left\{\phi_{\alpha_1^i}(x_1)\phi_{\alpha_2^i}(x_2)\cdots\phi_{\alpha_{N_1}^i}\left(x_{N_1}\right)\right\}\right], \tag{6}$$

where $\phi_{\alpha_l^i}(x_l)$; $l = 1, 2, 3, \cdots, N_1$ is a function belonging to the complete set of functions $\{\phi_i(x)\}$, and where the expression inside the square brackets represents the Slater determinant. Herein $P$ and $p$ are a permutation operator and the number of permutations in $P$, respectively, and the sum is performed over all permutations of $N_1$ electrons of system $S_1$, that is denoted as $P \in S_1$ in the sum notation. Equation (3) is a linear combination of Slater determinants with $a_{i,n}^{(1)}$ as expansion coefficients. Similarly, the eigenfunction for $H^{(2)}$ is given by

$$\Psi_m^{(2)}\left(y_1, y_2, \cdots, y_{N_2}\right) = \sum_j b_{j,m}^{(2)} \left[ \frac{1}{\sqrt{N_2\,!}} \sum_{P \in S_2} (-1)^p\, P \left\{ \psi_{\beta_1^j}(y_1)\psi_{\beta_2^j}(y_2)\cdots\psi_{\beta_{N_2}^j}\left(y_{N_2}\right) \right\} \right], \quad (7)$$

where $\psi_{\beta_l^j}(x_l);\ l = 1, 2, 3, \cdots, N_2$ is a function belonging to the complete set of functions $\{\psi_j(x)\}$.

Equation (7) is a linear combination of Slater determinants with $b_{j,m}^{(2)}$ as expansion coefficients.

***Antisymmetrizing operator***

The antisymmetrizing operator containing permutations of electrons within system $S_1$ is defined as

$$S_{N_1}^- = \sum_{P \in S_1} (-1)^p\, P. \quad (8)$$

Similarly, the antisymmetrizing operator containing permutations of electrons within system $S_2$ is defined as

$$S_{N_2}^- = \sum_{P \in S_2} (-1)^p\, P. \quad (9)$$

Considering the number of kinds of permutations, the operators $S_{N_1}^-$ and $S_{N_2}^-$ include $N_1!$ and $N_2!$ terms, respectively. The antisymmetrizing operator containing permutations of electrons within the composite system $S$ is defined as

$$S_N^- = \sum_{P \in S_1 + S_2} (-1)^p\, P. \quad (10)$$

It includes $N!$ terms. Here, we shall define the antisymmetrizing operator that includes at least one permutation of electrons between system $S_1$ and system $S_2$ such that

$$S'^- = \sum_{P \in \{S_1 \leftrightarrow S_2\}} (-1)^p\, P. \quad (11)$$

This operator includes $(N! - N_1!N_2!)$ terms. It is obvious that the antisymmetrizing operator $S_N^-$

can be decomposed into terms related to the permutations of electrons between system $S_1$ and system $S_2$ and terms not related to the permutations between system $S_1$ and system $S_2$. Namely, we have

$$\sum_{P\in S_1+S_2}(-1)^p P=\sum_{P\in\{S_1\leftrightarrow S_2\}}(-1)^p P+\left\{\sum_{P\in S_1}(-1)^p P\right\}\left\{\sum_{P\in S_2}(-1)^p P\right\}. \tag{12}$$

Using Eqs. (8), (9), (10) and (11), this relation can be rewritten as

$$S_N^- = S'^- + S_{N_1}^- S_{N_2}^- . \tag{13}$$

As will be shown in Sec. III, the condition $S'^- = 0$ is essential for satisfying the cluster property.

## III. Conditions for satisfying the cluster property

***Cluster property***

The cluster property is described by the following statement: *The expectation value of some physical quantity of a composite system is given by the sum or product of the expectation values of the individual systems constituting the composite system.*

This means that the individual systems are not entangled with each other. In this section we shall show conditions for satisfying the cluster property mentioned above.

***Indistinguishability of identical particles***

There exist electrons in system $S_1$ and system $S_2$, and of course, electrons are indistinguishable from one another. The indistinguishability of identical particles gives

$$PH^{(1)} = H^{(1)}P, \tag{14}$$

for some arbitrary permutation operator $P$ for electrons of system $S_1$. Similarly, for some arbitrary permutation operator $P$ for electrons of system $S_2$, we have

$$PH^{(2)} = H^{(2)}P. \tag{15}$$

Taking a summation on $P$ with multiplying the coefficient $(-1)^p$ in Eqs. (14) and (15), we get

$$S_{N_1}^- H^{(1)} = H^{(1)} S_{N_1}^- , \tag{16}$$

$$S_{N_2}^- H^{(2)} = H^{(2)} S_{N_2}^- , \tag{17}$$

respectively.

***Conditions for satisfying the cluster property***

Under the assumption that the number of electrons is conserved in each of system $S_1$ and system $S_2$, $\phi_{\alpha_l^i}(z);\ l = 1, 2, 3, \cdots, N_1$ has a nonzero probability amplitude for $z \in S_1$ in Eq. (6), and $\psi_{\beta_l^j}(z);\ l = 1, 2, 3, \cdots, N_2$ has a nonzero probability amplitude for $z \in S_2$ in Eq. (7). However, under this assumption, $\phi_{\alpha_l^i}(z)$ and $\psi_{\beta_l^j}(z)$ do not overlap, that is specifically described as

$$\phi_{\alpha_l^i}\left(y_j\right) = 0 \ \text{ for } \ l=1,2,3,\cdots,N_1,\ j=1,2,3,\cdots,N_2, \tag{18}$$

$$\psi_{\beta_l^j}\left(x_i\right) = 0 \ \text{ for } \ l=1,2,3,\cdots,N_2,\ i=1,2,3,\cdots,N_1 . \tag{19}$$

When Eqs. (18) and (19) hold, it is obvious that

$$S'^{-} = 0 \tag{20}$$

also holds. In this case, since electrons of system $S_1$ and those of system $S_2$ are completely isolated, they feel the different potentials, i.e., $v^{(1)} \neq v^{(2)}$, which is one of expressions or situations that are physically equivalent to Eq. (20). They are summarized in Appendix A.

In addition, we assume that the interaction between electrons of system $S_1$ and those of system $S_2$ is negligible. That is

$$W^{(12)}\left(x_i, y_j\right) = 0 . \tag{21}$$

Equations (20) and (21) should be assumed individually because the respective mechanisms for the emergence of entanglement are distinct. The entanglement related to $S'^{-} \neq 0$ is caused by the indistinguishability of identical particles, whereas the entanglement related to $W^{(12)}\left(x_i, y_j\right) \neq 0$ is caused by the interaction between particles. Under these two assumptions (20) and (21), the eigenfunction of a composite system $S$ can be expressed as a single tensor product of the eigenfunction of system $S_1$ and eigenfunction of system $S_2$. Specifically, we have

$$\Theta_{nm}\left(x_1, x_2, \cdots, x_{N_1}; y_1, y_2, \cdots, y_{N_2}\right) = \Phi_n^{(1)}\left(x_1, x_2, \cdots, x_{N_1}\right) \Psi_m^{(2)}\left(y_1, y_2, \cdots, y_{N_2}\right). \tag{22}$$

This can be simply proven, considering the fact that the eigenvalue equation for the Hamiltonian of system $S$ can be separated into eigenvalue equations for system $S_1$ and system $S_2$ by separation of variables. For reference, Eq. (22) is substantially antisymmetrized even within a composite system $S$.[29]

For simplicity, we shall denote eigenfunctions of a composite system $S$, system $S_1$ and system $S_2$ as $|\Theta_{nm}\rangle_{\mathrm{S}}$, $|\Phi_n^{(1)}\rangle_{\mathrm{S}_1}$, $|\Psi_m^{(2)}\rangle_{\mathrm{S}_2}$, respectively, by presenting the states without explicitly using coordinates. Then, Eq. (22) is expressed as

$$|\Theta_{nm}\rangle_{\mathrm{S}} = |\Phi_n^{(1)}\rangle_{\mathrm{S}_1} |\Psi_m^{(2)}\rangle_{\mathrm{S}_2}, \tag{23}$$

where the RHS is an abbreviation of the tensor product $|\Phi_n^{(1)}\rangle_{\mathrm{S}_1} \otimes |\Psi_m^{(2)}\rangle_{\mathrm{S}_2}$. Let $F$, $F^{(1)}$ and $F^{(2)}$ be the sums of single-particle operators over all particles belonging to a composite system $S$, system $S_1$ and system $S_2$, respectively, then we have

$$F = F^{(1)} + F^{(2)}. \tag{24}$$

Let us examine the specific expressions for $F^{(1)}$ and $F^{(2)}$. As for $F^{(1)}\left(F^{(2)}\right)$, we shall expand the field operator of electrons using the basis functions that have probability amplitudes within region of system $S_1$ $(S_2)$. Suppose that functions $U_i(x)$ have probability amplitudes only within region of system $S_1$ and that functions $V_j(y)$ have probability amplitudes only within region of system $S_2$. We define the following two types of field operator of electrons:

$$\hat{\psi}(x) = \sum_i \hat{C}_i U_i(x), \tag{25}$$

$$\hat{\psi}(y) = \sum_j \hat{D}_j V_j(y), \tag{26}$$

where $\hat{C}_i$ and $\hat{D}_j$ are annihilation operators to states $U_i(x)$ and $V_j(y)$, respectively. Here let us consider the sum of single-particle operators given by $\int dx \hat{\psi}^\dagger(x) f(x) \hat{\psi}(x)$, where $f(x)$ is a single-particle operator. Using Eq. (25), this operator can be rewritten as $\sum_i \sum_j \hat{C}_i^\dagger \hat{C}_j \int dx U_i^*(x) f(x) U_j(x)$, which acts on the many-electron state that contains non-zero occupation numbers for $\{U_i(x); i = 1, 2, \cdots\}$. Therefore, this operator is expressed as $F^{(1)}$. Furthermore, since Eq. (25) is an operator that annihilates an electron in region of system $S_1$, it is rewritten as $\hat{\psi}_{S_1}(x)$ or $\hat{\psi}(x);\ x \in S_1$. Namely, the sum of single-particle operators written using Eq. (25) can be expressed as

$$F^{(1)} = \int dx \hat{\psi}_{S_1}^\dagger(x) f(x) \hat{\psi}_{S_1}(x). \tag{27}$$

Similarly, Eq. (26) can be rewritten as $\hat{\psi}_{S_2}(y)$ or $\hat{\psi}(y);\ y \in S_2$. The sum of single-particle operators written using Eq. (26) can be expressed as

$$F^{(2)} = \int dx \hat{\psi}_{S_2}^\dagger(y) f(y) \hat{\psi}_{S_2}(y), \tag{28}$$

since it acts on the many-electron state that contains non-zero occupation numbers for $\{V_j(y);\ j = 1, 2, \cdots\}$.

As already mentioned, the wave function of a composite system $S$ is given by Eq. (23) under the conditions of Eqs. (20) and (21). The state $\left|\Phi_n^{(1)}\right\rangle_{S_1}$ contains non-zero occupation numbers for $\{U_i(x); i = 1, 2, \cdots\}$, but no information on the occupancy of $\{V_j(y);\ j = 1, 2, \cdots\}$, whereas the state $\left|\Psi_m^{(2)}\right\rangle_{S_2}$ contains non-zero occupation numbers for $\{V_j(y);\ j=1,2,\cdots\}$, but no information on the occupancy of $\{U_i(x);\ i=1,2,\cdots\}$. The operators $F^{(1)}$ and $F^{(2)}$ act only on $\left|\Phi_n^{(1)}\right\rangle_{S_1}$ and $\left|\Psi_m^{(2)}\right\rangle_{S_2}$, respectively. The expectation value of $F$ with respect to an eigenfunction of a composite system $S$ is given by

$$_S\langle\Theta_{nm}|F|\Theta_{nm}\rangle_S = {}_{S_1}\left\langle\Phi_n^{(1)}\right|F^{(1)}\left|\Phi_n^{(1)}\right\rangle_{S_1} + {}_{S_2}\left\langle\Psi_m^{(2)}\right|F^{(2)}\left|\Psi_m^{(2)}\right\rangle_{S_2}. \quad (29)$$

Here we use Eqs. (23) and (24), and normalization conditions of the eigenfunctions $\left|\Phi_n^{(1)}\right\rangle_{S_1}$ and $\left|\Psi_m^{(2)}\right\rangle_{S_2}$. Equation (29) indicates the presence of cluster property.

Let $G$, $G^{(1)}$ and $G^{(2)}$ be the sums of two-particle operators over all particles belonging to a composite system $S$, system $S_1$ and system $S_2$, respectively, and let $G^{(12)}$ be the sum of two-particle operators between particles in system $S_1$ and system $S_2$. Then we have the following relation:

$$G = G^{(1)} + G^{(2)} + G^{(12)}. \quad (30)$$

If the two-particle operator represents the interaction energy, the operator $G^{(12)}$ is trivially zero under the assumption of Eq. (21), i.e., $G^{(12)}=0$. At that time, the expectation value of $G$ with respect to an eigenfunction of a composite system $S$ is given by

$$_S\langle\Theta_{nm}|G|\Theta_{nm}\rangle_S = {}_{S_1}\left\langle\Phi_n^{(1)}\right|G^{(1)}\left|\Phi_n^{(1)}\right\rangle_{S_1} + {}_{S_2}\left\langle\Psi_m^{(2)}\right|G^{(2)}\left|\Psi_m^{(2)}\right\rangle_{S_2}. \quad (31)$$

Equation (31) also indicates the presence of cluster property. Of course, the operators $G^{(1)}$ and $G^{(2)}$ are written using the field operators $\hat{\psi}_{S_1}(x)$ or $\hat{\psi}(x);\ x\in S_1$ and $\hat{\psi}_{S_2}(y)$ or $\hat{\psi}(y);\ y\in S_2$, respectively, which are defined by Eqs. (25) and (26). The specific forms are given by

$$G^{(1)} = \frac{1}{2}\int dxdx'\hat{\psi}_{S_1}^\dagger(x)\hat{\psi}_{S_1}^\dagger(x')g(x,x')\hat{\psi}_{S_1}(x')\hat{\psi}_{S_1}(x), \quad (32)$$

$$G^{(2)}=\frac{1}{2}\int dydy'\hat{\psi}^{\dagger}_{S_2}(y)\hat{\psi}^{\dagger}_{S_2}(y')g(y,y')\hat{\psi}_{S_2}(y')\hat{\psi}_{S_2}(y), \tag{33}$$

respectively, where $g(x,x')$ is a two-particle operator. When deriving the cluster property Eq. (31), we use the fact that the operators $G^{(1)}$ and $G^{(2)}$ act only on $\left|\Phi_n^{(1)}\right\rangle_{S_1}$ and $\left|\Psi_m^{(2)}\right\rangle_{S_2}$, respectively.

In a similar way to Eqs. (29) and (31), the expectation values of operators of type of $F^{(1)}F^{(2)}\left(=F^{(1)}\otimes F^{(2)}\right)$ and $G^{(1)}G^{(2)}\left(=G^{(1)}\otimes G^{(2)}\right)$ for a composite system $S$ with respect to an eigenfunction of a composite system $S$ are given by

$${}_S\langle\Theta_{nm}|F^{(1)}F^{(2)}|\Theta_{nm}\rangle_S={}_{S_1}\left\langle\Phi_n^{(1)}\middle|F^{(1)}\middle|\Phi_n^{(1)}\right\rangle_{S_1}\cdot{}_{S_2}\left\langle\Psi_m^{(2)}\middle|F^{(2)}\middle|\Psi_m^{(2)}\right\rangle_{S_2}, \tag{34}$$

and

$${}_S\langle\Theta_{nm}|G^{(1)}G^{(2)}|\Theta_{nm}\rangle_S={}_{S_1}\left\langle\Phi_n^{(1)}\middle|G^{(1)}\middle|\Phi_n^{(1)}\right\rangle_{S_1}\cdot{}_{S_2}\left\langle\Psi_m^{(2)}\middle|G^{(2)}\middle|\Psi_m^{(2)}\right\rangle_{S_2}, \tag{35}$$

respectively. Equations (34) and (35) can also be regarded as cluster property. Thus, under the assumptions given by Eqs. (20) and (21), the cluster property holds true.

In addition, we shall consider more general operators as given below. Let $K$, $K^{(1)}$ and $K^{(2)}$ be the operators that contain one creation or annihilation operator for particles belonging to a composite system $S$, system $S_1$ and system $S_2$, respectively, and let $L$, $L^{(1)}$ and $L^{(2)}$ be the quadratic operators that contain two creation or annihilation operators for particles belonging to a composite system $S$, system $S_1$ and system $S_2$, respectively.

The operators $K^{(1)}$ and $K^{(2)}$ act only on the states of electrons in regions of system $S_1$ and system $S_2$, respectively. Therefore, in a similar way to Eq. (34), the expectation value of operator of type of $K^{(1)}K^{(2)}\left(=K^{(1)}\otimes K^{(2)}\right)$ for a composite system $S$ with respect to an eigenfunction of a composite system $S$ is given by

$${}_S\langle\Theta_{nm}|K^{(1)}K^{(2)}|\Theta_{nm}\rangle_S={}_{S_1}\left\langle\Phi_n^{(1)}\middle|K^{(1)}\middle|\Phi_n^{(1)}\right\rangle_{S_1}\cdot{}_{S_2}\left\langle\Psi_m^{(2)}\middle|K^{(2)}\middle|\Psi_m^{(2)}\right\rangle_{S_2}. \tag{36}$$

As examples of $K$, $K^{(1)}$ and $K^{(2)}$, we may give $\hat{\psi}(x);\ x\in S$, $\hat{\psi}(x');\ x'\in S^{(1)}$, $\hat{\psi}(x'');\ x''\in S^{(2)}$,

respectively. They correspond to the field operators that annihilate an electron within regions of a composite system $S$, system $S_1$ and system $S_2$, respectively.

The operators $L^{(1)}$ and $L^{(2)}$ act only on the states of electrons in regions of system $S_1$ and system $S_2$, respectively. Therefore, in a similar way to Eq. (35), the expectation value of operators of type of $L^{(1)}L^{(2)} \left(= L^{(1)} \otimes L^{(2)}\right)$ for a composite system $S$ with respect to an eigenfunction of a composite system $S$ is given by

$$ {}_S\langle \Theta_{nm} | L^{(1)} L^{(2)} | \Theta_{nm} \rangle_S = {}_{S_1}\langle \Phi_n^{(1)} | L^{(1)} | \Phi_n^{(1)} \rangle_{S_1} \cdot {}_{S_2}\langle \Psi_m^{(2)} | L^{(2)} | \Psi_m^{(2)} \rangle_{S_2} . \tag{37} $$

As examples of $L$, $L^{(1)}$ and $L^{(2)}$, we may give $\hat{\psi}(x)\hat{\psi}(X);\ x, X \in S$, $\hat{\psi}(x')\hat{\psi}(X');\ x', X' \in S^{(1)}$, $\hat{\psi}(x'')\hat{\psi}(X'');\ x'', X'' \in S^{(2)}$, respectively. They correspond to the product of the field operators that annihilate two electrons within regions of a composite system $S$, system $S_1$ and system $S_2$, respectively.

Equations (36) and (37) are also regarded as the cluster properties. However, it should be noted that types of operators ${}_{S_1}\langle \Phi_n^{(1)} | \hat{\psi}(x') | \Phi_n^{(1)} \rangle_{S_1}$, ${}_{S_2}\langle \Psi_m^{(2)} | \hat{\psi}(x'') | \Psi_m^{(2)} \rangle_{S_2}$, ${}_{S_1}\langle \Phi_n^{(1)} | \hat{\psi}(x')\hat{\psi}(X') | \Phi_n^{(1)} \rangle_{S_1}$ and ${}_{S_2}\langle \Psi_m^{(2)} | \hat{\psi}(x'')\hat{\psi}(X'') | \Psi_m^{(2)} \rangle_{S_2}$, which appear in Eqs. (36) and (37), are zero because the number of electrons is conserved in each of system $S_1$ and system $S_2$ from the assumption we made. The case where such types of operators remain nonzero is that of the superconducting system. It will be discussed in the subsequent section.

## IV. Cluster property in the superconducting state

In this section, we shall consider the cluster property in the superconducting state. To consider this issue, we prepare the following realistic situations for superconductors: both system $S_1$ and system $S_2$ consist of two systems that correspond to the superconducting electron system and normal electron system [30]. Specifically, system $S_1\,(S_2)$ consists of system $S_1^S\,\left(S_2^S\right)$ for the superconducting electrons and system $S_1^N\,\left(S_2^N\right)$ for the normal electrons.

There exist $N_1$ and $N_2$ electrons in system $S_1\left(= S_1^S + S_1^N\right)$ and $S_2\left(= S_2^S + S_2^N\right)$, respectively, and exist $N\,(= N_1 + N_2)$ electrons in a composite system $S$. Suppose that electrons pass between system $S_1^S\,\left(S_2^S\right)$ and system $S_1^N\,\left(S_2^N\right)$ but no electrons pass between system $S_1$ and $S_2$. In other words, the number of electrons is conserved in system $S_1$ and system $S_2$, respectively, but is fluctuating in system $S_1^S$ and system $S_2^S$, respectively (See Fig. 1). At that time, Eq. (20) still holds true. To reiterate, it is as follows:

$$S'^{-} = 0 \,. \tag{38}$$

Furthermore, we assume that the interaction between electrons of system $S_1$ and those of system $S_2$ is negligible. That is

$$W^{(12)}\left(x_i,\, y_j\right) = 0 \,. \tag{39}$$

In a similar way to the discussions done in the previous section, an eigenfunction of a composite system $S$ is given by a single tensor product of the eigenfunction of system $S_1$ and eigenfunction of system $S_2$. The same equation as Eq. (23) holds:

$$\left|\Theta_{nm}\right\rangle_{\mathrm{S}} = \left|\Phi_n^{(1)}\right\rangle_{\mathrm{S}_1}\left|\Psi_m^{(2)}\right\rangle_{\mathrm{S}_2}, \tag{40}$$

where the RHS is an abbreviation of the tensor product $\left|\Phi_n^{(1)}\right\rangle_{\mathrm{S}_1} \otimes \left|\Psi_m^{(2)}\right\rangle_{\mathrm{S}_2}$.

The eigenstate of system $S_1$ $(S_2)$ is expressed as a linear combination of the product states of system $S_1^N$ $\left(S_2^N\right)$ and system $S_1^S$ $\left(S_2^S\right)$ (Appendix B) [31]. This is a highly intuitive analogy, but the system $S_1^N$ $\left(S_2^N\right)$ and system $S_1^S$ $\left(S_2^S\right)$ correspond to the reservoir and the system in thermodynamics, respectively [31]. The number of electrons for the states of system $S_1^S$ $\left(S_2^S\right)$, which appear in such a linear combination of product states, is fluctuating.

Here, let us explicitly write down only the superconducting state in the eigenfunction of system $S_1$ $(S_2)$. Namely, we shall explicitly express only the superconducting state of system $S_1^S$ $\left(S_2^S\right)$ in eigenfunction $\left|\Phi_n^{(1)}\right\rangle_{\mathrm{S}_1}$ $\left(\left|\Psi_m^{(2)}\right\rangle_{\mathrm{S}_2}\right)$. The form of this notation is identical to that used for the BCS ground state [24-26]. The details of the state of system $S_1^S$ $\left(S_2^S\right)$ are described in Appendix B.

Using this notation, Eq. (40) reduces to

$$\left|\tilde{\Theta}_{nm}\right\rangle_{S_1^S+S_2^S}=\left|\tilde{\Phi}_n^{(1)}\right\rangle_{S_1^S}\left|\tilde{\Psi}_m^{(2)}\right\rangle_{S_2^S}, \tag{41}$$

where $\left|\tilde{\Phi}_n^{(1)}\right\rangle_{S_1^S}$ $\left(\left|\tilde{\Psi}_m^{(2)}\right\rangle_{S_2^S}\right)$ is the superconducting state extracted from within $\left|\Phi_n^{(1)}\right\rangle_{S_1}$ $\left(\left|\Psi_m^{(2)}\right\rangle_{S_2}\right)$, and where $\left|\tilde{\Theta}_{nm}\right\rangle_{S_1^S+S_2^S}$ is a single tensor product of these two superconducting states. It should be noted again that the numbers of electrons in superconducting states $\left|\tilde{\Phi}_n^{(1)}\right\rangle_{S_1^S}$ and $\left|\tilde{\Psi}_m^{(2)}\right\rangle_{S_2^S}$ fluctuate (see Appendix B).

We define system $S^S$, which consists of all superconducting electrons, as $S^S=S_1^S+S_2^S$. Let $F_S$, $F_{S_1^S}$ and $F_{S_2^S}$ be the sums of single-particle operators over all superconducting electrons belonging to systems $S^S$, system $S_1^S$ and system $S_2^S$, respectively. At that time, it is obvious to have

$$F_S=F_{S_1^S}+F_{S_2^S}. \tag{42}$$

In a similar way to discussions below Eq. (24), let us examine the specific expressions for $F_{S_1^S}$ and $F_{S_2^S}$. Let $u_i(x)$ $\left(v_j(y)\right)$ be the function that has probability amplitude only within region of system $S_1^S$ $\left(S_2^S\right)$. We define the following two types of field operator of superconducting electrons:

$$\hat{\psi}(x)=\sum_i \hat{c}_i u_i(x), \tag{43}$$

$$\hat{\psi}(y)=\sum_j \hat{d}_j v_j(y), \tag{44}$$

where $\hat{c}_i$ and $\hat{d}_j$ are annihilation operators to states $u_i(x)$ and $v_j(y)$, respectively. The sum of single-particle operators given by $\int dx\hat{\psi}^{\dagger}(x)f(x)\hat{\psi}(x)\left(=\sum_j \hat{c}_i\hat{c}_j\int dxu_i^*(x)f(x)u_j(x)\right)$ acts on the many-electron state that contains non-zero occupation numbers for $\{u_i(x); i=1,2,\cdots\}$. Therefore, this operator is expressed as $F_{S_1^S}$. Furthermore, since Eq. (43) is an operator that annihilates an electron in region of system $S_1^S$, it is rewritten as $\hat{\psi}_{S_1^S}(x)$ or $\hat{\psi}(x);\ x\in S_1^S$. Namely, the sum of single-particle operators written using Eq. (43) can be expressed as

$$F_{S_1^S}=\int dx\hat{\psi}^{\dagger}_{S_1^S}(x)f(x)\hat{\psi}_{S_1^S}(x)\,. \tag{45}$$

Similarly, Eq. (44) can be rewritten as $\hat{\psi}_{S_2^S}(y)$ or $\hat{\psi}(y);\ y\in S_2^S$. The sum of single-particle operators written using Eq. (44) can be expressed as

$$F_{S_2^S}=\int dx\hat{\psi}^{\dagger}_{S_2^S}(y)f(y)\hat{\psi}_{S_2^S}(y)\,. \tag{46}$$

The wave function of the entire superconducting electron system $S^S$ is given by Eq. (41) under the conditions of Eqs. (38) and (39). The superconducting state $\left|\tilde{\Phi}_n^{(1)}\right\rangle_{S_1^S}\left(\left|\tilde{\Psi}_m^{(2)}\right\rangle_{S_2^S}\right)$ contains non-zero occupation numbers for $\{u_i(x); i=1,2,\cdots\}\left(\{v_j(y); j=1,2,\cdots\}\right)$, but no information on the occupancy of $\{v_j(y); j=1,2,\cdots\}$ $(\{u_i(x); i=1,2,\cdots\})$. The operators $F_{S_1^S}$ and $F_{S_2^S}$ act only on $\left|\tilde{\Phi}_n^{(1)}\right\rangle_{S_1^S}$ and $\left|\tilde{\Psi}_m^{(2)}\right\rangle_{S_2^S}$, respectively. The expectation value of Eq. (42) with respect to the wave function of the entire superconducting electron system $S^S$ is calculated as

$${}_{S_1^S+S_2^S}\left\langle\tilde{\Theta}_{nm}\left|F_S\right|\tilde{\Theta}_{nm}\right\rangle_{S_1^S+S_2^S}={}_{S_1^S}\left\langle\tilde{\Phi}_n^{(1)}\left|F_{S_1^S}\right|\tilde{\Phi}_n^{(1)}\right\rangle_{S_1^S}+{}_{S_2^S}\left\langle\tilde{\Psi}_m^{(2)}\left|F_{S_2^S}\right|\tilde{\Psi}_m^{(2)}\right\rangle_{S_2^S}\,. \tag{47}$$

Here we use normalization conditions of the eigenfunctions $\left|\tilde{\Phi}_n^{(1)}\right\rangle_{S_1^S}$ and $\left|\tilde{\Psi}_m^{(2)}\right\rangle_{S_2^S}$. Equation (47) indicates the presence of cluster property in the superconducting state.

Next, let $G_S$, $G_{S_1^S}$ and $G_{S_2^S}$ be the sums of two-particle operators over all superconducting electrons belonging to systems $S^S$, system $S_1^S$ and system $S_2^S$, respectively, and let $G_{S_1^S S_2^S}$ be the sum of two-particle operators between particles in system $S_1^S$ and those in system $S_2^S$. Then, we have the following relation:

$$G_S = G_{S_1^S} + G_{S_2^S} + G_{S_1^S S_2^S} \,. \tag{48}$$

If the two-particle operator represents the interaction energy, the operator $G_{S_1^S S_2^S}$ is trivially zero under the assumption of Eq. (39), i.e., $G_{S_1^S S_2^S} = 0$. At that time, the expectation value of $G_S$ with respect to an eigenfunction of a composite system $S^S$ is given by

$${}_{\mathrm{S}_1^S+\mathrm{S}_2^S}\left\langle \tilde{\Theta}_{nm} \right| G_S \left| \tilde{\Theta}_{nm} \right\rangle_{\mathrm{S}_1^S+\mathrm{S}_2^S} = {}_{\mathrm{S}_1^S}\left\langle \tilde{\Phi}_n^{(1)} \right| G_{S_1^S} \left| \tilde{\Phi}_n^{(1)} \right\rangle_{\mathrm{S}_1^S} + {}_{\mathrm{S}_2^S}\left\langle \tilde{\Psi}_m^{(2)} \right| G_{S_2^S} \left| \tilde{\Psi}_m^{(2)} \right\rangle_{\mathrm{S}_2^S} . \tag{49}$$

Equation (49) also indicates the presence of cluster property in the superconducting state. The operators $G_{S_1^S}$ and $G_{S_2^S}$ are written using the field operators $\hat{\psi}_{S_1^S}(x)$ or $\hat{\psi}(x)$; $x \in S_1^S$ and $\hat{\psi}_{S_2^S}(y)$ or $\hat{\psi}(y)$; $y \in S_2^S$, respectively. The specific forms are given by

$$G_{S_1^S} = \frac{1}{2}\int dxdx' \hat{\psi}_{S_1^S}^{\dagger}(x)\hat{\psi}_{S_1^S}^{\dagger}(x') g(x,x') \hat{\psi}_{S_1^S}(x')\hat{\psi}_{S_1^S}(x) , \tag{50}$$

$$G_{S_2^S} = \frac{1}{2}\int dydy' \hat{\psi}_{S_2^S}^{\dagger}(y)\hat{\psi}_{S_2^S}^{\dagger}(y') g(y,y') \hat{\psi}_{S_2^S}(y')\hat{\psi}_{S_2^S}(y) , \tag{51}$$

respectively, where $g(x,x')$ is a two-particle operator. When deriving the cluster property Eq. (49), we use the fact that the operators $G_{S_1^S}$ and $G_{S_2^S}$ act only on $\left|\tilde{\Phi}_n^{(1)}\right\rangle_{\mathrm{S}_1^S}$ and $\left|\tilde{\Psi}_m^{(2)}\right\rangle_{\mathrm{S}_2^S}$, respectively.

In a similar way to Eqs. (47) and (49), the expectation values of operators of type of $F_{S_1^S} F_{S_2^S} \left(= F_{S_1^S} \otimes F_{S_2^S}\right)$ and $G_{S_1^S} G_{S_2^S} \left(= G_{S_1^S} \otimes G_{S_2^S}\right)$ for a composite system $S^S$ with respect to an eigenfunction of a composite system $S^S$ are given by

$$ {}_{\mathrm{S}_1^S+\mathrm{S}_2^S}\left\langle \tilde{\Theta}_{nm} \middle| F_{S_1^S} F_{S_2^S} \middle| \tilde{\Theta}_{nm} \right\rangle_{\mathrm{S}_1^S+\mathrm{S}_2^S} = {}_{\mathrm{S}_1^S}\left\langle \tilde{\Phi}_n^{(1)} \middle| F_{S_1^S} \middle| \tilde{\Phi}_n^{(1)} \right\rangle_{\mathrm{S}_1^S} \cdot {}_{\mathrm{S}_2^S}\left\langle \tilde{\Psi}_m^{(2)} \middle| F_{S_2^S} \middle| \tilde{\Psi}_m^{(2)} \right\rangle_{\mathrm{S}_2^S} , \tag{52} $$

and

$$ {}_{\mathrm{S}_1^S+\mathrm{S}_2^S}\left\langle \tilde{\Theta}_{nm} \middle| G_{S_1^S} G_{S_2^S} \middle| \tilde{\Theta}_{nm} \right\rangle_{\mathrm{S}_1^S+\mathrm{S}_2^S} = {}_{\mathrm{S}_1^S}\left\langle \tilde{\Phi}_n^{(1)} \middle| G_{S_1^S} \middle| \tilde{\Phi}_n^{(1)} \right\rangle_{\mathrm{S}_1^S} \cdot {}_{\mathrm{S}_2^S}\left\langle \tilde{\Psi}_m^{(2)} \middle| G_{S_2^S} \middle| \tilde{\Psi}_m^{(2)} \right\rangle_{\mathrm{S}_2^S} , \tag{53} $$

respectively. Equations (52) and (53) can also be regarded as the cluster property. Thus, under the assumptions given by Eqs. (38) and (39), the cluster property holds in the superconducting state.

In addition, we shall consider more general operators shown below. Let $K_S$, $K_{S_1^S}$ and $K_{S_2^S}$ be the operators that contain one creation or annihilation operator for superconducting electrons belonging to a composite system $S^S$, system $S_1^S$ and system $S_2^S$, respectively, and let $L_S$, $L_{S_1^S}$ and $L_{S_2^S}$ be the quadratic operators that contain two creation or annihilation operators for superconducting electrons belonging to a composite system $S^S$, system $S_1^S$ and system $S_2^S$, respectively.

The operators $K_{S_1^S}$ and $K_{S_2^S}$ act only on the states of superconducting electrons in regions of system $S_1^S$ and system $S_2^S$, respectively. Therefore, in a similar way to Eq. (36), the expectation value of operator of type of $K_{S_1^S} K_{S_2^S} \left(= K_{S_1^S} \otimes K_{S_2^S}\right)$ for a composite system $S^S$ with respect to an eigenfunction of a composite system $S^S$ is given by

$$ {}_{\mathrm{S}_1^S+\mathrm{S}_2^S}\left\langle \tilde{\Theta}_{nm} \middle| K_{S_1^S} K_{S_2^S} \middle| \tilde{\Theta}_{nm} \right\rangle_{\mathrm{S}_1^S+\mathrm{S}_2^S} = {}_{\mathrm{S}_1^S}\left\langle \tilde{\Phi}_n^{(1)} \middle| K_{S_1^S} \middle| \tilde{\Phi}_n^{(1)} \right\rangle_{\mathrm{S}_1^S} \cdot {}_{\mathrm{S}_2^S}\left\langle \tilde{\Psi}_m^{(2)} \middle| K_{S_2^S} \middle| \tilde{\Psi}_m^{(2)} \right\rangle_{\mathrm{S}_2^S} . \tag{54} $$

As examples of $K_S$, $K_{S_1^S}$ and $K_{S_2^S}$, we may give $\hat{\psi}(x);\ x \in S^S$, $\hat{\psi}(x');\ x' \in S_1^S$, $\hat{\psi}(x'');\ x'' \in S_2^S$, respectively. They correspond to the field operators that annihilate a superconducting electron within regions of a composite system $S^S$, system $S_1^S$ and system $S_2^S$, respectively.

The operators $L_{S_1^S}$ and $L_{S_2^S}$ act only on the states of superconducting electrons in regions of system $S_1^S$ and system $S_2^S$, respectively. Therefore, in a similar way to Eq. (37), the expectation value of operators of type of $L_{S_1^S} L_{S_2^S} \left(= L_{S_1^S} \otimes L_{S_2^S}\right)$ for a composite system $S^S$ with respect to an eigenfunction of a composite system $S^S$ is given by

$$ {}_{S_1^S+S_2^S}\left\langle \tilde{\Theta}_{nm} \middle| L_{S_1^S} L_{S_2^S} \middle| \tilde{\Theta}_{nm} \right\rangle_{S_1^S+S_2^S} = {}_{S_1^S}\left\langle \tilde{\Phi}_n^{(1)} \middle| L_{S_1^S} \middle| \tilde{\Phi}_n^{(1)} \right\rangle_{S_1^S} \cdot {}_{S_2^S}\left\langle \tilde{\Psi}_m^{(2)} \middle| L_{S_2^S} \middle| \tilde{\Psi}_m^{(2)} \right\rangle_{S_2^S} . \tag{55} $$

As examples of $L_S$, $L_{S_1^S}$ and $L_{S_2^S}$, we may give $\hat{\psi}(x)\hat{\psi}(X);\ x, X \in S^S$, $\hat{\psi}(x')\hat{\psi}(X');\ x', X' \in S_1^S$, $\hat{\psi}(x'')\hat{\psi}(X'');\ x'', X'' \in S_2^S$, respectively. They correspond to the product of the field operators that annihilate two superconducting electrons within regions of a composite system $S^S$, system $S_1^S$ and system $S_2^S$, respectively. Equations (54) and (55) also exhibit the cluster properties in the superconducting state.

***A specific form of the cluster property in the superconducting state***

As mentioned above Eqs. (45) and (46), the field operators for superconducting electrons can be denoted as follows:

$$ \hat{\psi}_{S_1^S}(x) := \hat{\psi}(x);\ x \in S_1^S, \tag{56} $$

$$ \hat{\psi}_{S_1^S}(X) := \hat{\psi}(X);\ X \in S_1^S, \tag{57} $$

$$ \hat{\psi}_{S_2^S}(y) := \hat{\psi}(y);\ y \in S_2^S, \tag{58} $$

$$ \hat{\psi}_{S_2^S}(Y) := \hat{\psi}(Y);\ Y \in S_2^S\ . \tag{59} $$

Using the field operators with these notations, we define the following operators

$$ L_{S_1^S}(x, X) = \hat{\psi}^\dagger_{S_1^S}(x)\hat{\psi}^\dagger_{S_1^S}(X), \tag{60} $$

$$ L_{S_2^S}(y, Y) = \hat{\psi}_{S_2^S}(y)\hat{\psi}_{S_2^S}(Y), \tag{61} $$

where the regions of $x, X$ and $y, Y$ are, of course, given by $x, X \in S_1^S$ and $y, Y \in S_2^S$,

respectively. As examples of $L_{S_1^S}$ and $L_{S_2^S}$ in Eq. (55), we shall take Eqs. (60) and (61), respectively. At that time, using Eqs. (60) and (61) instead of $L_{S_1^S}$ and $L_{S_2^S}$, Eq. (55) is expressed as

$$
\begin{aligned}
&{}_{\mathrm{S}_1^S+\mathrm{S}_2^S}\left\langle \tilde{\Theta}_{nm} \middle| \hat{\psi}^\dagger_{S_1^S}(x)\hat{\psi}^\dagger_{S_1^S}(X)\hat{\psi}_{S_2^S}(y)\hat{\psi}_{S_2^S}(Y) \middle| \tilde{\Theta}_{nm} \right\rangle_{\mathrm{S}_1^S+\mathrm{S}_2^S} \\
&= {}_{\mathrm{S}_1^S}\left\langle \tilde{\Phi}^{(1)}_n \middle| \hat{\psi}^\dagger_{S_1^S}(x)\hat{\psi}^\dagger_{S_1^S}(X) \middle| \tilde{\Phi}^{(1)}_n \right\rangle_{\mathrm{S}_1^S} \cdot {}_{\mathrm{S}_2^S}\left\langle \tilde{\Psi}^{(2)}_m \middle| \hat{\psi}_{S_2^S}(y)\hat{\psi}_{S_2^S}(Y) \middle| \tilde{\Psi}^{(2)}_m \right\rangle_{\mathrm{S}_2^S}
\end{aligned}
$$

with

$$x, X \in S_1^S, \text{ and } y, Y \in S_2^S. \tag{62}$$

This is also regarded as the cluster property in the superconducting state. Using the normalization conditions of the eigenfunctions $\left|\tilde{\Phi}^{(1)}_n\right\rangle_{\mathrm{S}_1^S}$ and $\left|\tilde{\Psi}^{(2)}_m\right\rangle_{\mathrm{S}_2^S}$, Eq. (62) can be rewritten as

$$
\begin{aligned}
&{}_{\mathrm{S}_1^S+\mathrm{S}_2^S}\left\langle \tilde{\Theta}_{nm} \middle| \hat{\psi}^\dagger_{S_1^S}(x)\hat{\psi}^\dagger_{S_1^S}(X)\hat{\psi}_{S_2^S}(y)\hat{\psi}_{S_2^S}(Y) \middle| \tilde{\Theta}_{nm} \right\rangle_{\mathrm{S}_1^S+\mathrm{S}_2^S} \\
&= {}_{\mathrm{S}_1^S+\mathrm{S}_2^S}\left\langle \tilde{\Theta}_{nm} \middle| \hat{\psi}^\dagger_{S_1^S}(x)\hat{\psi}^\dagger_{S_1^S}(X) \middle| \tilde{\Theta}_{nm} \right\rangle_{\mathrm{S}_1^S+\mathrm{S}_2^S} \cdot {}_{\mathrm{S}_1^S+\mathrm{S}_2^S}\left\langle \tilde{\Theta}_{nm} \middle| \hat{\psi}_{S_2^S}(y)\hat{\psi}_{S_2^S}(Y) \middle| \tilde{\Theta}_{nm} \right\rangle_{\mathrm{S}_1^S+\mathrm{S}_2^S}
\end{aligned}
$$

with

$$x, X \in S_1^S, \text{ and } y, Y \in S_2^S. \tag{63}$$

This is another expression for the cluster property in the superconducting state.

## V. Cluster property appearing when considering the order parameter of the superconducting state

The superconducting state has been defined using the Bose-Einstein condensation (BEC) of the fermion system [4, 5, 19, 20]. First, we shall confirm the definition of the superconducting state. The operator of the second-order reduced density matrix (RDM2) is given by [4, 32]

$$\hat{\rho}^{(2)}(x_1, x_2; x_1', x_2') = \frac{1}{2}\hat{\psi}^\dagger(x_1')\hat{\psi}^\dagger(x_2')\hat{\psi}(x_2)\hat{\psi}(x_1), \tag{64}$$

where $\hat{\psi}(x)$ and $\hat{\psi}^\dagger(x)$ are field operators of electrons. The RDM2, which is uniquely determined by the many-electron state $|\Psi\rangle$, is defined as

$$\rho_\Psi^{(2)}(x_1,x_2;x_1',x_2') = \langle\Psi|\frac{1}{2}\hat{\psi}^\dagger(x_1')\hat{\psi}^\dagger(x_2')\hat{\psi}(x_2)\hat{\psi}(x_1)|\Psi\rangle\,. \tag{65}$$

The eigenvalue equation for $\rho_\Psi^{(2)}(x_1,x_2;x_1',x_2')$ is given by

$$\sum_{\zeta_1'}\sum_{\zeta_2'}\int \rho_\Psi^{(2)}(x_1,x_2;x_1',x_2')\nu_\Psi(x_1',x_2')\,d^3r_1'd^3r_2' = n_{\nu_\Psi}^{(2)}\nu_\Psi(x_1,x_2)\,, \tag{66}$$

where the eigenfunction $\nu_\Psi(x_1,x_2)$ is a two-particle state that is often called geminal [4, 32], and where the eigenvalue $n_{\nu_\Psi}^{(2)}$ corresponds to the occupation number of $\nu_\Psi(x_1,x_2)$ within $|\Psi\rangle$ when the many-electron state is interpreted as the set of two-particle states [4, 32]. The spectral decomposition of $\rho_\Psi^{(2)}(x_1,x_2;x_1',x_2')$ is written as

$$\rho_\Psi^{(2)}(x_1,x_2;x_1',x_2') = \sum_{\nu_\Psi} n_{\nu_\Psi}^{(2)}\nu_\Psi(x_1,x_2)\nu_\Psi^*(x_1',x_2')\,. \tag{67}$$

***Definition of the superconducting state***

The geminal $\nu_\Psi(x_1,x_2)$ can be rewritten by using the relative coordinates $\boldsymbol{\rho}$ and coordinates of the center of gravity $\boldsymbol{R}$. Let it be denoted as $\nu_\Psi(\boldsymbol{R},\boldsymbol{\rho};\zeta_1,\zeta_2)$. There exist two types of geminals, depending on whether there is an attractive force between the constituent particles [4]. They have been as follows: *Type I* corresponds to the geminal, in which there is no attractive force between constituent particles, and which is not localized in terms of both $\boldsymbol{R}$ and $\boldsymbol{\rho}$. *Type II* corresponds to the geminal, in which there is some attractive force between constituent particles, and which is localized in terms of $\boldsymbol{\rho}$ but is not in terms of $\boldsymbol{R}$ [4].

The many-electron state is in the superconducting state when the following condition is satisfied: *The occupation number for a certain geminal is* $O(N)$*, and such a geminal belongs to type II* [4, 19, 20]. This geminal has been named as "maximum geminal" in the previous papers [4, 27, 33], and hereafter we will use this name. This is the definition of the superconducting state on the basis of the BEC of the fermion system.

***Order parameter of the superconducting state***

According to the definition of the superconducting state, we shall derive the OPSS. Let us split Eq. (67) into the term with the maximum eigenvalue and the other ones. Namely, we have

$$\rho_{\Psi}^{(2)}(x_1,x_2;x_1',x_2') = n_{\nu_{\Psi}^{\max}}^{(2)} \nu_{\Psi}^{\max}(x_1,x_2)\nu_{\Psi}^{\max *}(x_1',x_2') + \sum_{\nu_{\Psi}\left(\neq \nu_{\Psi}^{\max}\right)} n_{\nu_{\Psi}}^{(2)} \nu_{\Psi}(x_1,x_2)\nu_{\Psi}^{*}(x_1',x_2'). \quad (68)$$

According to the definition of the superconducting state, if the many-electron state $|\Psi\rangle$ is in the superconducting state, the following conditions are satisfied:

$$\begin{cases} n_{\nu_{\Psi}^{\max}}^{(2)} = O(N),\ \nu_{\Psi}^{\max}(x_1,x_2) \text{ is of type II}, \\ n_{\nu_{\Psi}}^{(2)} = O(1),\ \nu_{\Psi}(x_1,x_2) \text{ is of type I for } \nu_{\Psi}\left(\neq \nu_{\Psi}^{\max}\right). \end{cases} \quad (69)$$

Here we consider the limit operation given by

$$\begin{cases} \boldsymbol{r}_1 \approx \boldsymbol{r}_2,\ \boldsymbol{r}_1' \approx \boldsymbol{r}_2', \\ \quad\text{and} \\ |\boldsymbol{r}_1 - \boldsymbol{r}_1'| \to \infty. \end{cases} \quad (70)$$

This limit means that relative distance between two particles that constitute the geminal becomes small and that distance between each geminal is made infinite. It can be shown that only the first term of RHS of Eq. (68) remains nonzero in the limit of Eq. (70) if $|\Psi\rangle$ fulfills the conditions of Eq. (69) [4, 20]. Namely, if $|\Psi\rangle$ is in the superconducting state, only the term of the maximum geminal in the spectral decomposition of $\rho_{\Psi}^{(2)}(x_1,x_2;x_1',x_2')$ remains nonzero in the limit of Eq. (70). We have

$$\rho_{\Psi}^{(2)}(x_1,x_2;x_1',x_2') \underset{\text{limit of Eq.(56)}}{\longrightarrow} n_{\nu_{\Psi}^{\max}}^{(2)} \nu_{\Psi}^{\max}(x_1,x_2)\nu_{\Psi}^{\max *}(x_1',x_2'). \quad (71)$$

On the other hand, we shall apply the limit of Eq. (70) to Eq. (65). Let us consider two systems consisting of superconducting electrons, system $S_1^S$ and system $S_2^S$, and let a composite system be system $S^S$. Furthermore, let $x_1, x_2$ and $x_1', x_2'$ be the coordinates of system $S_1^S$ and system $S_2^S$, respectively. That is

$$\begin{cases} x_1 \in S_1^S,\ x_2 \in S_1^S, \\ x_1' \in S_2^S,\ x_2' \in S_2^S. \end{cases} \quad (72)$$

Here we consider the limit given by Eq. (70). This limit contains $|\boldsymbol{r}_1 - \boldsymbol{r}_1'| \to \infty$, which means that system $S_1^S$ and system $S_2^S$ are significantly separated from each other. At that time, the interaction between particles belonging to system $S_1^S$ and system $S_2^S$ seems to be substantially zero. In addition, the overlap between wave functions of particles belonging to system $S_1^S$ and system $S_2^S$ seems to be also zero. That is, Eqs. (38) and (39) hold true in the limit of Eq. (70). As a result, the cluster property in the superconducting state, i.e., Eq. (63) is satisfied in the limit of Eq. (70). In this case we have

$$
\begin{aligned}
&\frac{1}{2}\,{}_S\langle \Psi | \hat{\psi}^{\dagger}_{S_2^S}(x_1')\hat{\psi}^{\dagger}_{S_2^S}(x_2')\hat{\psi}_{S_1^S}(x_2)\hat{\psi}_{S_1^S}(x_1) | \Psi \rangle_S \\
&\xrightarrow[\text{limit of Eq.(56)}]{} \frac{1}{2}\,{}_S\langle \Psi | \hat{\psi}^{\dagger}_{S_2^S}(x_1')\hat{\psi}^{\dagger}_{S_2^S}(x_2') | \Psi \rangle_S \cdot {}_S\langle \Psi | \hat{\psi}_{S_1^S}(x_2)\hat{\psi}_{S_1^S}(x_1) | \Psi \rangle_S \\
&\qquad = \frac{1}{2}\,{}_S\langle \Psi | \hat{\psi}_{S_1^S}(x_1)\hat{\psi}_{S_1^S}(x_2) | \Psi \rangle_S \cdot {}_S\langle \Psi | \hat{\psi}^{\dagger}_{S_2^S}(x_2')\hat{\psi}^{\dagger}_{S_2^S}(x_1') | \Psi \rangle_S .
\end{aligned}
\tag{73}
$$

Comparing Eq. (71) and Eq. (73), we get the following relation:

$$
{}_S\langle \Psi | \hat{\psi}_{S_1^S}(x_1)\hat{\psi}_{S_1^S}(x_2) | \Psi \rangle_S = \sqrt{2 n^{(2)}_{\nu_\Psi^{\max}}}\, \nu_\Psi^{\max}(x_1, x_2) . \tag{74}
$$

Thus, it is concluded that the OPSS is given by Eq. (74).

## VI. Entanglement property between the constituent particles of the maximum geminal

In the previous section, it is shown that the cluster property holds in the superconducting state when $S'^{-} = 0$ and $W^{(12)} = 0$. It should be noted that this cluster property holds between the maximum geminals, not between the constituent particles of the maximum geminal.

In this section we shall discuss the cluster property and entanglement property between the constituent particles of the maximum geminal. The cluster property is broken between the constituent particles of the maximum geminal due to $S'^{-} \neq 0$ and/or $W^{(12)} \neq 0$, and the entanglement between them emerges. In this case, system $S_1$ and system $S_2$ correspond to two constituent particles of the maximum geminal in the same way as the previous paper [33].

The explicit form of the maximum geminal of the BCS ground state is written as [27]

$$
\nu^{\max}_{\Theta_{\text{BCS}}}(\boldsymbol{r}_1, \boldsymbol{r}_2; \zeta_1, \zeta_2) = \Phi_{\text{singlet}}(\zeta_1, \zeta_2) \sum_{\boldsymbol{k}} A_{\boldsymbol{k}} e^{i\boldsymbol{k}\cdot\boldsymbol{\rho}} , \tag{75}
$$

where $\boldsymbol{\rho}$ is the relative coordinates and where $\Phi_{\text{singlet}}(\zeta_1,\zeta_2)$ is the spin singlet state given by

$$\Phi_{\text{singlet}}(\zeta_1,\zeta_2)=\frac{1}{\sqrt{2}}\{\chi_\uparrow(\zeta_1)\chi_\downarrow(\zeta_2)-\chi_\downarrow(\zeta_1)\chi_\uparrow(\zeta_2)\}. \tag{76}$$

Using the free electron energy measured from the chemical potential $\mu$ and the gap parameter $\Delta_0$, the coefficient $A_{\boldsymbol{k}}$ in Eq. (75) is given by

$$A_{\boldsymbol{k}}=\frac{1}{V}\frac{1}{\sqrt{\pi N(\mu)}}\left\{\frac{\Delta_0}{\left(\frac{\hbar^2k^2}{2m}-\mu\right)^2+\Delta_0^2}\right\}^{\frac{1}{2}}e^{\mathrm{i}\theta}, \tag{77}$$

where $N(\mu)$ is the density of states at the chemical potential and where $e^{\mathrm{i}\theta}$ and $V$ are an arbitrary phase factor and volume of the system, respectively [27]. As the gap parameter $\Delta_0$ increases, the coefficient $A_{\boldsymbol{k}}$ becomes more uniformly and extensively distributed. That is, the larger $\Delta_0$ is, the larger the entanglement entropy becomes [33].

On the other hand, according to the BCS theory [25, 26], the gap parameter is given by

$$\Delta_0=2\hbar\omega_D e^{-\frac{1}{V_0N(\mu)}}, \tag{78}$$

where $\omega_D$ is the Debye frequency and where $V_0$ is the attractive interaction energy between electrons in a Cooper pair. If system $S_1$ and system $S_2$ correspond to two constituent particles of the maximum geminal, respectively, the physical quantity which is related to $W^{(12)}$ is an attractive interaction energy between electrons in a maximum geminal, i.e., $V_0$. From Eq. (78), the larger the attractive interaction energy $V_0$ is, or equivalently the larger the effect of $W^{(12)}$ is, the larger the gap parameter $\Delta_0$ becomes, thereby increasing entanglement entropy as well.

In addition, it is safe to assume that the physical quantity which is related to $S'^{-}$ is the mean distance between electrons composing the maximum geminal. It has been derived in the previous work [27], and is explicitly expressed as

$$\bar{\rho}=\frac{1}{2\sqrt{2}}\frac{\hbar v_{\text{F}}}{\Delta_0}, \tag{79}$$

where $v_{\mathrm{F}}$ is the Fermi velocity. From Eq. (79), the smaller the mean distance $\bar{\rho}$ is, or equivalently the larger the effect of $S'^{-}$ is, the larger the gap parameter $\Delta_0$ becomes, resulting in a larger entanglement entropy.

From these results, the following can be concluded for the maximum geminal of the BCS ground state: As the degree of the deviation of $W^{(12)}$ from zero and/or that of $S'^{-}$ increase, that is, as $V_0$ increases and/or $\bar{\rho}$ decreases, the entanglement entropy increases as well. This seems to be a reasonable result from the viewpoint of the conditions for satisfying the cluster property.

## VII. Concluding remarks

The cluster property holds true only upon the fulfillment of the conditions given by Eqs. (20) and (21) or Eqs. (38) and (39) for the superconducting state. Equivalently, of course, we can say that the entanglement between systems does not occur without a cause. Such a cause is expressed as $S'^{-} \neq 0$ and/or $W^{(12)} \neq 0$, as already shown in this paper.

In order to understand this better, let us consider the following hypothetical example. Suppose that the relations such that $S'^{-} = 0$ and $W^{(12)} = 0$ hold between system $S_1$ and system $S_2$. Further suppose that the ground state is doubly degenerate in both systems. Specifically, we suppose that

$$\begin{cases} H^{(1)} |\Phi_1\rangle_{S_1} = \varepsilon_G^{(1)} |\Phi_1\rangle_{S_1}, \\ H^{(1)} |\Phi_2\rangle_{S_1} = \varepsilon_G^{(1)} |\Phi_2\rangle_{S_1}, \end{cases} \tag{80}$$

and

$$\begin{cases} H^{(2)} |\Psi_1\rangle_{S_2} = \varepsilon_G^{(2)} |\Psi_1\rangle_{S_2}, \\ H^{(2)} |\Psi_2\rangle_{S_2} = \varepsilon_G^{(2)} |\Psi_2\rangle_{S_2}. \end{cases} \tag{81}$$

Since the cluster property holds under the assumptions of $S'^{-} = 0$ and $W^{(12)} = 0$, the ground state of a composite system $S$ composed of system $S_1$ and system $S_2$ is expressed in the form of a single tensor product of the eigenfunctions of system $S_1$ and system $S_2$. Namely we have

$$|\Theta\rangle_S = \left\{ a_1 |\Phi_1\rangle_{S_1} + a_2 |\Phi_2\rangle_{S_1} \right\} \left\{ b_1 |\Psi_1\rangle_{S_2} + b_2 |\Psi_2\rangle_{S_2} \right\}, \tag{82}$$

where $a_1, a_2, b_1, b_2$ are complex constants. The state $|\Theta\rangle_S$ is a ground state for the Hamiltonian $H = H^{(1)} + H^{(2)}$, and the corresponding eigenvalue is $\varepsilon_G = \varepsilon_G^{(1)} + \varepsilon_G^{(2)}$. The state $|\Theta\rangle_S$ is a pure state composed of a single tensor product term, and is therefore not entangled. It is quite natural because both $S'^{-}$ and $W^{(12)}$, which cause the entanglement, are zero.

Solely from the viewpoint of energy, the following state is, in principle, a possible ground state of a composite system $S$:

$$|\Xi\rangle_S = C\left\{|\Phi_1\rangle_{S_1}|\Psi_1\rangle_{S_2} + |\Phi_2\rangle_{S_1}|\Psi_2\rangle_{S_2}\right\}, \tag{83}$$

where $C$ is the normalization constant. It is obvious that Eq. (83) is an entangled eigenstate for $H$, and its corresponding eigenvalue is $\varepsilon_G = \varepsilon_G^{(1)} + \varepsilon_G^{(2)}$. However, this state is actually excluded as a ground state by the fact that there is no cause for entanglement in this case. It should be stressed again that the entanglement between systems does not occur without a cause given by $S'^{-} \neq 0$ and/or $W^{(12)} \neq 0$.

**Acknowledgement**

This work was partially supported by Grant-in-Aid for Scientific Research from the Japan Society for the Promotion of Science and MEXT KAKENHI Grant (Nos. 23K03250, 25K07164, 24H01618 and 26H00607).

**Appendix A**: *Expressions or situations physically equivalent to* $S'^{-} = 0$

As mentioned in the text, we assume that the number of electrons is conserved in each of two systems, system $S_1$ and system $S_2$. This assumption can be restated in the following ways:

(1) Since the number of particles belonging to each individual system, system $S_1$ and system $S_2$, is conserved, the overlap of wave functions of particles belonging to each individual system can be considered negligible.

(2) The statement (1) means that the potential energy that is felt by particles belonging to each individual system is different. Namely, we have

$$v^{(1)}(z) \neq v^{(2)}(z), \tag{A1}$$

where $v^{(1)}(x_j)$ and $v^{(2)}(y_j)$ are potential energies that are contained in $h^{(1)}\left(x_j, \nabla_{x_j}\right)$ and $h^{(2)}\left(y_j, \nabla_{y_j}\right)$, respectively.

(3) The statement (1) also implies that Eqs. (18) and (19) hold true, which directly yields

$$S'^{-} = 0 . \tag{A2}$$

(4) Using the permutation operator $P$, which includes at least one permutation of electrons between system $S_1$ and system $S_2$, the statement (2) can be expressed as follows:

$$P\left(H^{(1)} + H^{(2)} + W^{(12)}\right) \neq \left(H^{(1)} + H^{(2)} + W^{(12)}\right)P . \tag{A3}$$

(5) The statement (4) means that there is no need to impose antisymmetry requirements on the wave function of a composite system $S$.

**Appendix B:** *Superconducting state of system* $S_1$ *or system* $S_2$

First, let us consider a state $|\Psi\rangle_{S_1}^{2,N_1-2}$ for system $S_1$, which consists of a superconducting state with two particles and a normal state with $N_1 - 2$ particles. It can be generally written as

$$|\Psi\rangle_{S_1}^{2,N_1-2} = \sum_{a,b} \sum_{i,j,k,\cdots} \alpha_{ab,ijk\cdots} |ab\rangle_{S_1^S} |ijk\cdots\rangle_{S_1^N} , \tag{B1}$$

where $|ab\rangle_{S_1^S}$ is a superconducting state that is generated using the creation operators for superconducting electrons, and where $|ijk\cdots\rangle_{S_1^N}$ is a normal state that is generated using the creation operators for normal electrons. Similarly, a state $|\Psi\rangle_{S_1}^{4,N_1-4}$ $\left(|\Psi\rangle_{S_1}^{6,N_1-6}, |\Psi\rangle_{S_1}^{8,N_1-8}, \cdots\right)$ for system $S_1$, which consists of a superconducting state with four (six, eight, $\cdots$) particles and a normal state with $N_1 - 4$ $(N_1 - 6,\ N_1 - 8, \cdots)$ particles are generally written as

$$\begin{aligned}
|\Psi\rangle_{S_1}^{4,N_1-4} &= \sum_{a,b,c,d} \sum_{i,j,k,\cdots} \beta_{abcd,ijk\cdots} |abcd\rangle_{S_1^S} |ijk\cdots\rangle_{S_1^N} , \\
|\Psi\rangle_{S_1}^{6,N_1-6} &= \sum_{a,b,c,d,e,f} \sum_{i,j,k,\cdots} \gamma_{abcdef,ijk\cdots} |abcdef\rangle_{S_1^S} |ijk\cdots\rangle_{S_1^N} , \\
|\Psi\rangle_{S_1}^{8,N_1-8} &= \sum_{a,b,c,d,e,f,g,h} \sum_{i,j,k,\cdots} \delta_{abcdefgh,ijk\cdots} |abcdefgh\rangle_{S_1^S} |ijk\cdots\rangle_{S_1^N} , \\
&\cdots\cdots, \\
&\cdots\cdots.
\end{aligned} \tag{B2}$$

The eigenfunction of system $S_1$ is generally given by a linear combination of the above states. Namely, we have

$$\begin{aligned}\left|\Phi_n^{(1)}\right\rangle_{S_1} &= A\sum_{i,j,k,\cdots}\eta_{ijk\cdots}\left|ijk\cdots\right\rangle_{S_1^N} \\ &+B\sum_{a,b}\sum_{i,j,k,\cdots}\alpha_{ab,ijk\cdots}\left|ab\right\rangle_{S_1^S}\left|ijk\cdots\right\rangle_{S_1^N} \\ &+C\sum_{a,b,c,d}\sum_{i,j,k,\cdots}\beta_{abcd,ijk\cdots}\left|abcd\right\rangle_{S_1^S}\left|ijk\cdots\right\rangle_{S_1^N} \\ &+D\sum_{a,b,c,d,e,f}\sum_{i,j,k,\cdots}\gamma_{abcdef,ijk\cdots}\left|abcdef\right\rangle_{S_1^S}\left|ijk\cdots\right\rangle_{S_1^N} \\ &+E\sum_{a,b,c,d,e,f,g,h}\sum_{i,j,k,\cdots}\delta_{abcdefgh,ijk\cdots}\left|abcdefgh\right\rangle_{S_1^S}\left|ijk\cdots\right\rangle_{S_1^N} \\ &+\cdots\cdots,\end{aligned} \tag{B3}$$

where the first term of RHS denotes the state with no superconducting particles, and where $A, B, C, D, E, \cdots$ are expansion coefficients.

If only the superconducting states are explicitly written down from Eq. (B3), we obtain the superconducting state which is analogous to the BCS ground state and exhibits a fluctuating particle number.

**References**

[1] W. Kohn, E. K. U. Gross, and L. N. Oliveira, J. de Physique (Paris) **50**, 2601 (1989).

[2] L. N. Oliveira, E. K. U. Gross, and W. Kohn, Phys. Rev. Lett. **60**, 2430 (1988).

[3] R. M. Dreizler and E. K. U. Gross, *Density Functional Theory* (Springer-Verlag, Berlin, 1990) Chap. 3.9.

[4] K. Higuchi, E. Miki and M. Higuchi, J. Phys. Soc. Jpn. **86**, 064704 (2017).

[5] M. Ueda, *Fundamentals and New Frontiers of Bose-Einstein Condensation* (World Scientific, Singapore, 2010) Chap. 1.

[6] K. Higuchi, H. Niwa and M. Higuchi, J. Phys. Soc. Jpn. **86**, 104705 (2017).

[7] K. Higuchi, N. Matsumoto, Y. Kamijo and M. Higuchi, Phys. Rev. B **102**, 014515 (2020).

[8] K. Higuchi, N. Matsumoto, Y. Kamijo and M. Higuchi, J. Phys.: Condens. Matter **33**, 435602 (2021).

[9] K. Higuchi and M. Higuchi, Phys. Rev. B **106,** 184504 (2022).

[10] M. Lüders, M. A. L. Marques, N. N. Lathiotakis, A. Floris, G. Profeta, L. Fast, A. Continenza, S. Massidda, and E. K. U. Gross, Phys. Rev. B **72**, 024545 (2005).

[11] M. A. L. Marques, M. Lüders, N. N. Lathiotakis, Gianni Profeta, A. Floris, L. Fast, A. Continenza, E. K. U. Gross, and S. Massidda, Phys. Rev. B **72**, 024546 (2005).

[12] A. Sanna, G. Profeta, A. Floris, A. Marini, E. K. U. Gross, and S. Massidda, Phys. Rev. B **75**,

020511 (2007).
[13] J. A. Flores-Livas, M. Amsler, T. J. Lenosky, L. Lehtovaara, S. Botti, M. A. L. Marques, and S. Goedecker, Phys. Rev. Lett. **108**, 117004 (2012).
[14] J. Schmidt, C. L. Benavides-Riveros, and M. A. L. Marques, Phys. Rev. B **99**, 024502 (2019).
[15] A. Sanna, C. Pellegrini, and E. K. U. Gross, Phys. Rev. Lett. **125**, 057001 (2020).
[16] C. Pellegrini and A. Sanna, Nat. Rev. Phys. **6**, 509 (2024).
[17] Z. Wan, G-H. Zhong, R. Zhang, and H-Q. Lin, Phys. Rev. B **112**, L220504 (2025).
[18] X. Fan, P. Jing, C. Zhang, J. Wang, C. Ma, S. Gong, C. Zhao, T. Wang, and Y. An, Phys. Rev. B **113**, 214521 (2026).
[19] C. N. Yang, Rev. Mod. Phys. **34**, 694 (1962).
[20] A. J. Leggett, *Quantum Liquids* (Oxford University Press, New York, 2006) Chap. 2.
[21] A. J. Leggett, J. Phys. Colloques **41**, C7-19 (1980).
[22] A. J. Leggett, in *Modern Trends in the Theory of Condensed Matter*, edited by A. Pękalski and J. Przystawa, Lect. Notes Phys. **115** (Springer-Verlag, Berlin, 1980), 13–27.
[23] Chap. 5 of Ref. [5].
[24] J. Bardeen, L. N. Cooper and J. R. Schrieffer, Phys. Rev. **108**, 1175 (1957).
[25] See, for example, M. Tinkham, *Introduction to Superconductivity* (Dover Pub., New York, 1996) Chap. 3.
[26] See, for example, P. G. de Gennes, *Superconductivity of Metals and Alloys* (W. A. Benjamin, New York, 1966) Chap. 4.
[27] K. Higuchi and M. Higuchi, J. Phys. Commun. **5**, 095003 (2021).
[28] M. A. Nielsen and I. L. Chuang, *Quantum Computation and Quantum Information* (Cambridge University Press, New York, 2000) Chap. 11.3.
[29] A. Messiah, Quantum Mechanics (Dover Pub. Inc., New York, 1999) Chap. 14.8.
[30] We know the fact that two types of electrons, namely superconducting electrons and normal electrons, coexist in the superconductor [11][12]. Normal electrons play a role of the reservoir to the system consisting of superconducting electrons. As a result, the number of superconducting electrons fluctuates and simultaneously that of the normal electrons also fluctuates.
[31] W. Greiner, L. Neise and H. Stroecker, *Thermodynamics and Statistical Mechanics* (Springer-Verlag, New York, 1995) Chap. 10.3.
[32] R. G. Parr and W. Yang, *Density-Functional Theory of Atoms and Molecules* (Oxford University Press, New York, 1989) Chap. 2.
[33] K. Higuchi, I. Tanno, R. Ito and M. Higuchi, J. Phys. Commun. **8**, 045003 (2024).

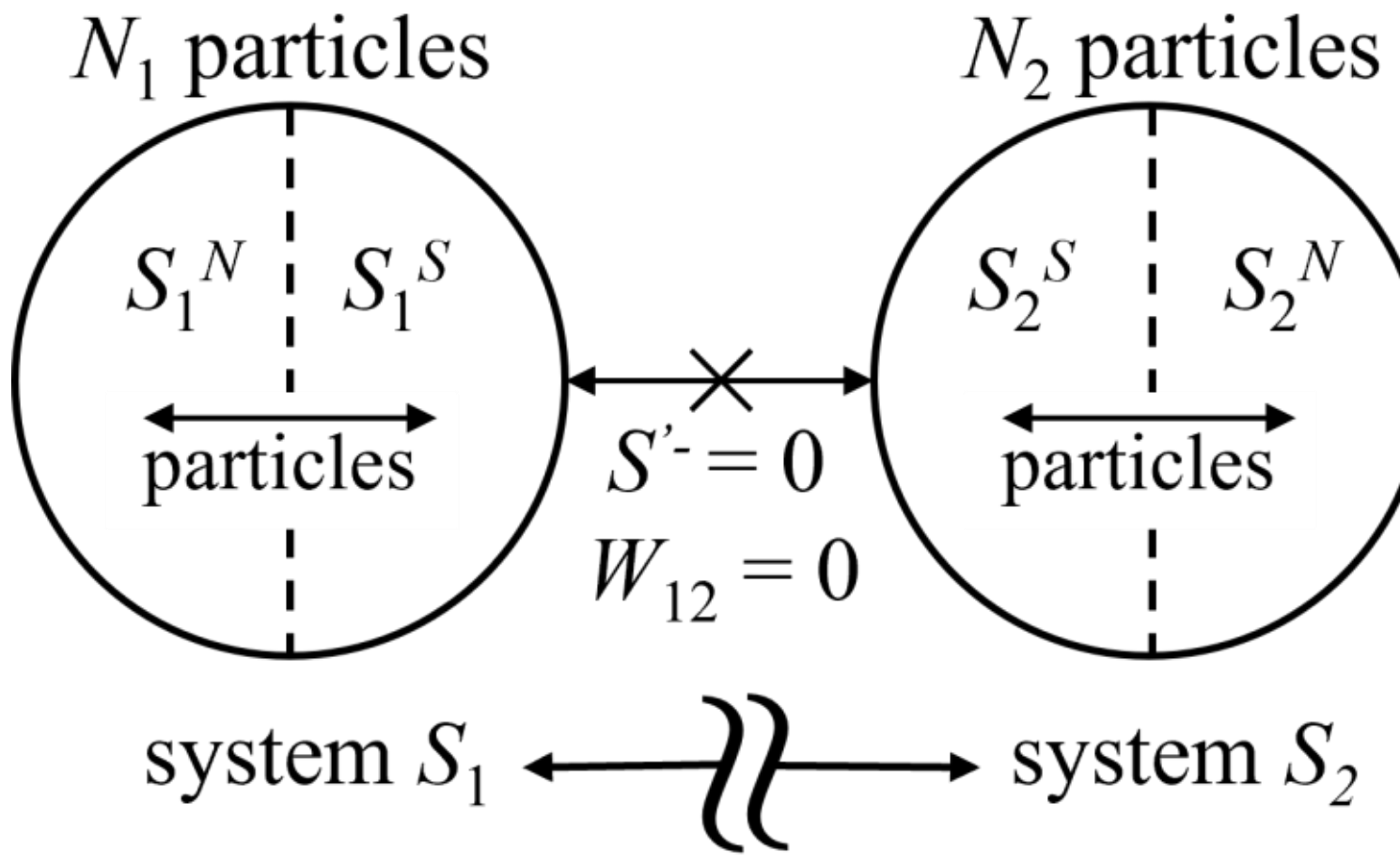

Fig. 1: Schematic explanation for the superconducting systems.